\documentclass[aps,prl,twocolumn,showpacs,superscriptaddress,floatfix]{revtex4-1}
\usepackage{graphicx,graphics}
\usepackage{dcolumn}
\usepackage{amsmath,amssymb,amsfonts}
\usepackage{latexsym,verbatim}
\usepackage{bm}
\usepackage{color}
\usepackage[breaklinks=true,colorlinks,citecolor=blue,linkcolor=blue,urlcolor=blue]{hyperref}

\def\be{\begin{equation}}
\def\ee{\end{equation}}

\newcommand{\beq}    {\begin{equation}}
\newcommand{\enq}    {\end{equation}}
\newcommand{\ceq}[1] {(\ref{#1})}

\newcommand{\eps}    {\epsilon}
\newcommand{\kk}     {{\bm k}}

\newcommand{\rr}     {{\bm r}}
\newcommand{\RR}     {{\bm R}}
\newcommand{\ssb}   {{\bm S}}

\newcommand{\nimp}   {n_{\rm imp}}

\newcommand{\tk}    {T_{\rm K}}

\begin{document}
\title{Kondo effect and non-Fermi liquid behavior in Dirac and Weyl semimetals}

\author{Alessandro Principi}
\email{principia@missouri.edu}
\affiliation{Department of Physics and Astronomy, University of Missouri, Columbia, Missouri 65211, USA}	
\affiliation{Institute for Molecules and Materials, Radboud University Nijmegen, Heijndaalseweg 135, 6525 AJ, Nijmegen, The Netherlands}
\author{Giovanni Vignale}
\affiliation{Department of Physics and Astronomy, University of Missouri, Columbia, Missouri 65211, USA}
\author{E. Rossi}
\affiliation{Department of Physics, College of William and Mary, Williamsburg, VA 23187, USA}

\begin{abstract}
We study the Kondo effect in three-dimensional (3D) Dirac materials and Weyl semimetals.
We find the scaling of the Kondo temperature with respect to the doping $n$ and the coupling 
$J$ between the moment of the magnetic impurity and the carriers of the semimetal.
We consider the interplay of long-range scalar disorder and Kondo 
screening and find that it  causes the Kondo effect to be
characterized not by a Kondo temperature but by a distribution of
Kondo temperatures with features that cause the appearance of strong non-Fermi liquid behavior. 
We then consider the effect of Kondo screening, and of the interplay of Kondo screening
and long-range scalar disorder, on the transport properties of Weyl semimetals.
Finally we compare the properties of the Kondo effect in 3D and two-dimensional (2D)
Dirac materials like graphene and topological insulators.
\end{abstract}

\pacs{65.80.Ck,72.20.Pa,72.80.Vp}
\maketitle

In Weyl and Dirac semimetals (SMs)~\cite{Volovik_book,Fradkin_prb_1986,Wan_prb_2011,Burkov_prb_2011,Wang_prb_2013,Liu_science_2014,Neupane_natcomm_2014} the conduction and valence bands touch at isolated points of the Brillouin zone (BZ) named ``Weyl nodes'' in Weyl SMs and ``Dirac points'' (DPs) in Dirac SMs. 
Around these points the electronic excitations behave as three-dimensional (3D) massless
Dirac fermions characterized by a density-independent Fermi velocity $v_{\rm F}$. 
Weyl SMs are expected to exhibit unique properties~\cite{Adler_pr_1969,Fukushima_prd_2008,Kharzeev_progPNP_2014} and to have surface states forming ``Fermi arcs''~\cite{Burkov_prb_2011,Zyuzin_prb_2012,Zyuzin_prb_2012_2,Wan_prb_2011,Imura_prb_2011,Hosur_prb_2012,Halasz_prb_2012,Okugawa_prb_2014,Haldane_arxiv_2014,Potter_NatComm_2014}.
The eigenstates of the bare Hamiltonian are non-degenerate in the case of Weyl SMs~\cite{Wan_prb_2011,Burkov_prb_2011,Volovik_book,Wang_prb_2013}.
Conversely, in Dirac SMs the eigenstates are doubly degenerate, {\it i.e.} each Dirac point corresponds to two copies of overlapping Weyl nodes with opposite chiralities~\cite{Wang_prb_2012}.
The linear dispersion around the nodes is expected to give rise to anomalous transport properties in both 3D Dirac and Weyl SMs~\cite{Burkov_prb_2011,Lundgren_prb_2014}.
Graphene~\cite{CastroNeto_rmp_2009,DasSarma_rmp_2011,Kotov_rmp_2012} and the surface states of 3D topological insulators (TIs)~\cite{Hasan_rmp_2010,Qi_rmp_2011} constitute the two-dimensional (2D) counterpart of 3D SMs~\cite{Hasan_rmp_2010,Qi_rmp_2011}. 

At low temperature, magnetic impurities strongly affect the properties of any electron liquid.
The ``Kondo effect''~\cite{Kondo_PTP_1964,Hewson_book} is characterized by a temperature scale $\tk$: when the temperature ($T$) is larger than $\tk$ the electrons of the host material are only weakly scattered by the impurity; for $T < T_{\rm K}$ the (antiferromagnetic) coupling grows non-perturbatively and leads to the formation of a many-body singlet with the electron liquid, which completely screens the impurity magnetic moment.

In this work we show that the unique band structure of 3D Dirac and Weyl SMs strongly affects the nature of the Kondo effect in these systems. We 
(i)   obtain the dependence of $\tk$ on the doping level
      of the SM and on the strength of the antiferromagnetic electron-impurity coupling $J$, 
(ii)  show that the interplay of linear dispersion around the nodes, Kondo effect, and long-range scalar disorder induces a strong 
      non-Fermi liquid (NFL) behavior~\cite{Dobrosavljevic_prl_1992,Miranda_JPCM_1996,Miranda_prl_1997,Miranda_RPP_2005} in these systems,
(iii) obtain the effect of the Kondo screening, and of the interplay of Kondo screening
      and long-range scalar disorder, on the transport properties and on the magnetic susceptibility of SMs.
      These quantities can be used to address experimentally the Kondo effect and the NFL behavior.
Finally we present
a systematic comparison of the properties of the Kondo effect
between 3D and 2D Dirac SMs~\cite{Withoff_prl_1990,Cassanello_prb_1996,Ingersent_prb_1996,Vojta_prb_2004,Fritz_prb_2004,Sengupta_prb_2008,Cornaglia_prl_2009,Vojta_epl_2010,Zitko_prb_2010,Wehling_prb_2010,Fritz_rpp_2013,Orignac_prb_2013,Miranda_prb_2014,Mastrogiuseppe_prb_2014}.

In Dirac and Weyl SMs the low-energy states around one of the DPs are described by the Hamiltonian 
$H_0 = {\hat c}^\dagger_{\kk\sigma}(\hbar v_F \kk\cdot{\bm\tau}_{\sigma\sigma'} -\mu\delta_{\sigma\sigma'}) {\hat c}_{\kk\sigma'}$,
where ${\hat c}^\dagger_{\kk\sigma}$ (${\hat c}_{\kk\sigma}$) creates (annihilates) an electron with momentum $\kk$ and spin (or pseudospin) $\sigma$, and $\mu$ is the chemical potential. Hereafter we set $\hbar=1$. 
For TIs and Weyl SMs (graphene and 3D Dirac SMs) ${\bm\tau}_{\sigma\sigma'}$ is the vector formed by the $2\times 2$ Pauli matrices in spin (pseudospin) space. 
The contribution of Fermi arcs to the Kondo effect in Weyl SMs is negligible, since it requires a flip of the electron spin, and consequently a jump between different surfaces of the SM.
Thus, the differences between Weyl and Dirac SMs, besides the extra spin degeneracy $g_s = 2$ of Dirac eigenstates, turn out to be inessential for our purposes.

In the presence of diluted magnetic impurities the
system is described by the Hamiltonian
$
 H = H_0 + H_J
$
where
$
 H_J= J \sum_{\rr,\RR} {\hat c}^\dagger_{\rr\sigma}{\bm\tau}_{\sigma\sigma'}{\hat c}_{\rr\sigma'}\cdot\ssb\delta(\rr-\RR),
$
with $\ssb$ the magnetic moment of the impurities, $\{\RR\}$ their
positions, and $J$ the strength of the (antiferromagnetic) coupling between the impurities and the carriers.
To treat this term we use a large-N
expansion~\cite{Read_JPCSSP_1983,Bickers_rmp_1987},
by which $\ssb$ is expressed in terms of auxiliary creation (annihilation) fermionic operators
${\hat f}^\dagger_\sigma$ (${\hat f}_\sigma$)
satisfying the constraint
$n_f=\sum_\sigma {\hat f}^\dagger_\sigma {\hat f}_{\sigma}=1$,
with  $\sigma = 1,\ldots, N_\sigma$.
We set $N_\sigma=2$ at the end of the calculation, which corresponds to the case of a magnetic impurity with $|\ssb|=1/2$.
In terms of the ${\hat f}$-operators the coupling term $H_J$ takes the form $H_J = J \sum_{{\bm k},{\bm k}',\sigma} {\hat c}^\dagger_{{\bm k}\sigma}{\hat c}_{{\bm k}'\sigma'}{\hat f}^\dagger_{\sigma'}{\hat f}_\sigma$. 

The large-N expansion allows a mean field treatment of the Kondo problem~\cite{Read_JPCSSP_1983}, and is  known to return accurate and reliable results for the case of diluted magnetic impurities~\cite{Read_JPCSSP_1983,Bickers_rmp_1987,Rossi_prl_2006}.
We decouple the quartic interaction term $H_J$ via a Hubbard-Stratonovich field $s\sim \sum_{{\bm k},\sigma} \langle  {\hat f}^\dagger_{\sigma} {\hat c}_{{\bm k}\sigma} \rangle$, which describes the hybridization between ``localized'' (${\hat f}$) and ``itinerant'' (${\hat c}$) electronic states.  The constraint $n_f=1$ is enforced by the introduction of a Lagrange multiplier $\mu_f$, which plays the role of the chemical potential of the $f$-electrons.
Approximating $s$ and $\mu_f$ as static (mean-)fields, we finally obtain the effective action
\begin{equation} \label{eq:effective_action_2}
{\cal S}_{\rm eff} \!\! = 
\!\!
\frac{1}{k_{\rm B} T}
\!\!
\left[
\frac{2}{\pi} \int d\varepsilon~
\frac{\arctan \left[\frac{\pi}{2} \frac{|s|^2 {\cal N}(\varepsilon+\mu)}{\varepsilon - \mu_f} \right]}{e^{\varepsilon/(k_B T)} + 1}
+ \frac{|s|^2}{J} - \mu_f 
\right]
~,
\end{equation}
where the integral is bound between $-D-\mu$ and $D-\mu$, 
$D$ is a cut-off corresponding to half the bandwidth of the SM,
${\cal N}(\varepsilon) = V N_{\rm w} \varepsilon^2/(2\pi^2 \hbar^3 v_{\rm F}^3)$ is the density of states (DOS) with
$N_{\rm w}$ the number of DPs, and $V$ the volume of the system.
The corresponding ${\cal S}_{\rm eff}$
for the 2D case is obtained by replacing
${\cal N}(\varepsilon) \to V N_{\rm w} |\varepsilon|/(2\pi \hbar^2 v_{\rm F}^2)$. By minimizing $S_{\rm eff}$ within the saddle point approximation~\cite{Bickers_rmp_1987}
we obtain the self-consistent equations for  $|s|^2$ and $\mu_f$.

We identify $\tk$ as the highest temperature for which
the self-consistent equations have a non-trivial solution. 
Depending on the value of $\mu$ we can have two distinct situations. For $\mu=0$, i.e. when the chemical potential of the 3D SM lies exactly at the DP,
we obtain
\beq \label{eq:tk_mu_0}
T_{\rm K} = D \frac{\sqrt{3}}{\pi}\sqrt{1 - \frac{2}{{\cal N}(D) J}}~, \hspace{0.7cm} \mu=0
~.
\enq
Eq.~(\ref{eq:tk_mu_0}) is valid only for $J$ larger than the critical value $J_{\rm cr}=2/{\cal N}(D)$; $T_{\rm K}$ vanishes when this condition is not met. A similar result is found in 2D, for which one obtains $\tk=D~\big[1-1/({\cal N}(D)J)\big]/\ln(4)$~\cite{Sengupta_prb_2008,Orignac_prb_2013,Withoff_prl_1990,Cassanello_prb_1996,Gonzalez_prb_1998}. 
In the 2D case $J_{\rm cr} = 1/{\cal N}(D)$. 
Numerical renormalization group (NRG) calculations~\cite{Ingersent_prb_1996} show that, for ${\cal N}(\varepsilon) \sim |\varepsilon|^a$ (with $a>1/2$) and in the presence of
perfect particle-hole symmetry, the Kondo effect cannot be realized for any value of $J$.
The previous results should be intended to describe 2D and 3D SMs close to, but not exactly at, a particle-hole symmetric situation.
This is likely the most realistic condition given that  
in real systems typically there is no particle-hole symmetry.
Local fluctuations take the local $\mu$ away from the Dirac point almost everywhere in the sample.
Moreover, in many systems like graphene and TIs (in 2D) or the Weyl SM ${\rm Cd_2As_3}$~\cite{Liu_natmat_2014} (in 3D)
the Fermi velocities of the conduction and valence bands are different.

\begin{figure}[t]
\begin{center}
\begin{tabular}{c}
\includegraphics[width=0.48\columnwidth]{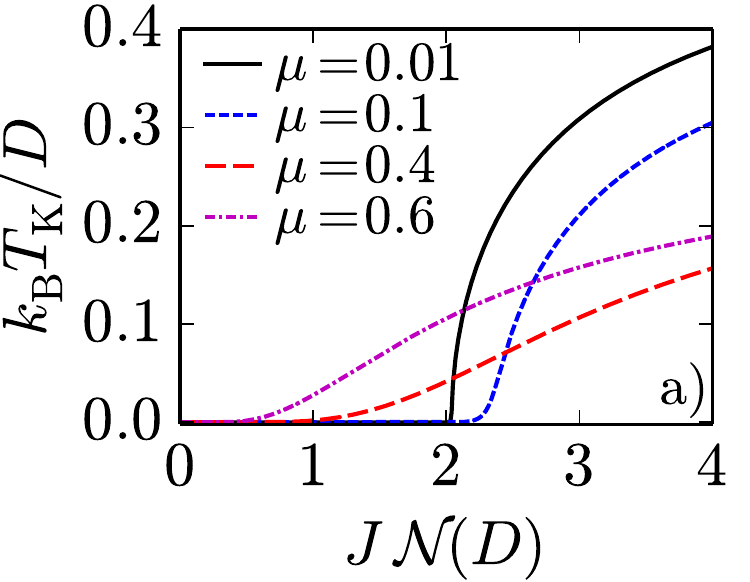}
\includegraphics[width=0.48\columnwidth]{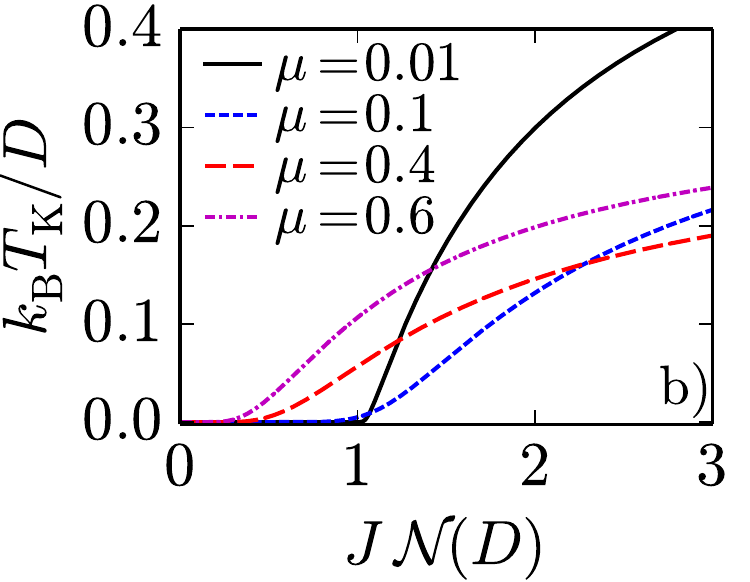}
\end{tabular}
\end{center}
\caption{(Color online).
 $\tk$ as a function of $\hat J=J{\cal N}(D)$ for different values of $\mu/D$  for a 3D, a), and 2D, b) Dirac SM.
\label{fig:one}}
\end{figure}

When $\mu \neq 0$, in the limit $k_{\rm B} T_{\rm K} \ll \mu \ll D$ and $J\lesssim J_{\rm cr}$ we obtain
\beq \label{eq:tk_mu_not_0}
\tk = D \exp\left[\frac{1 - 2 /(J{\cal N}(D))}{2 \mu^2/D^2} \right]~, \hspace{0.7cm} \mu\neq 0
~.
\enq
%
In 2D~\cite{Vojta_epl_2010} and for $J\lesssim J_{\rm cr}$ we have instead $\tk = \kappa(\mu) e^{ [1 - 1/({\cal N}(D)J)]/|\mu/D| }$, where $\kappa(\mu)=\mu^2/D$ [$\kappa(\mu)=D$] for $\mu>0$ [$\mu<0$]. 
For $J\gtrsim J_{\rm cr}$, $\tk$ can be obtained numerically.
Fig.~\ref{fig:one} shows $\tk$ for 3D and 2D SMs as a function of $J$ (both smaller and larger than $J_{\rm c}$) and for different values of $\mu>0$.

We now investigate the effect of long-range scalar disorder on the Kondo effect.
In Dirac SMs, differently from ``standard'' metals, charged impurities induce~\cite{Nomura_prl_2006,Ando_JPSJ_2006,Galitski_prb_2007} long-range carrier density inhomogeneities~\cite{Adam_pnas_2007,Rossi_prl_2008}. Such inhomogeneities have been observed in direct imaging experiments in graphene~\cite{Li_natphys_2008,Zhang_natphys_2009,Deshpande_prb_2009} and TIs~\cite{Hasan_rmp_2010,Beidenkopf_natphys_2011,Kim_natphys_2012}. Since the DOS of 3D Dirac SMs scales with the density, as $\sim n^{2/3}$ in 3D and $\sim n^{1/2}$ in 2D, the long-range fluctuations of the carrier density reflect on the DOS and therefore on 
$\tk$, Eq.~\ceq{eq:tk_mu_not_0}.
The Kondo effect is not characterized anymore by a single value of $\tk$, but by a distribution of Kondo temperatures $P(T_{\rm K})$~\cite{Miranda_prb_2014}. A similar situation was predicted to occur in metals close to a metal-insulator transition (MIT)~\cite{Dobrosavljevic_prl_1992}. 

We consider a Gaussian density distribution
$P_n(n)$ centered around the average doping $\bar n$, with 
standard deviation $\sigma_n$ (proportional to the number of dopants), {\it i.e.}
$P_n(n) = \exp\big[ -(n-{\bar n})^2/(2\sigma_n^2) \big]/(\sqrt{2\pi} \sigma_n)$. This assumption for $P_n(n)$ has been shown to be well justified for the case of 2D graphene \cite{Chen_natphys_2008,Adam_ssc_2009,Rossi_prb_2009} and we expect it to be a reasonable model also for 3D SMs. Using this expression for $P_n(n)$ and the fact that $\mu\sim n^{1/3}$, from Eq.~\ceq{eq:tk_mu_not_0} we obtain 
\begin{equation} \label{eq:TK_distribution_3D}
P^{({\rm 3D})} (T_{\rm K}) = \frac{3 D^3 T_{\rm K}^{-1}}{8\sqrt{\pi} \sigma_\mu^3} \sqrt{ \frac{\big(1 - J_{\rm c}/J\big)^3}{\ln^5(k_{\rm B} T_{\rm K}/D)} }
\!\! \sum_{\lambda=\pm 1} e^{ - \frac{(\mu^3-\lambda {\bar \mu}^3)^2}{2\sigma_\mu^6} } 
~,
\end{equation}
where ${\bar \mu} = v_{\rm F} (6\pi^2 {\bar n}/N_{\rm w})^{1/3}$, $\sigma_\mu = v_{\rm F} (6\pi^2 \sigma_n/N_{\rm w})^{1/3}$, and $\mu\equiv \mu(T_K)$ is obtained by inverting Eq.~(\ref{eq:tk_mu_not_0}). In so doing, we neglected the change of the local DOS due to the scalar part of the potential of the magnetic impurity, which is significant only when $\mu\sim 0$. Due to the strong carrier density inhomogeneities induced by the long-range disorder, even when ${\bar n}=0$ the area of the sample where $\mu \sim 0$ has measure zero. Therefore, the change of the DOS can be neglected.

\begin{figure}[t]
\begin{center}
\begin{tabular}{c}
\includegraphics[width=0.48\columnwidth]{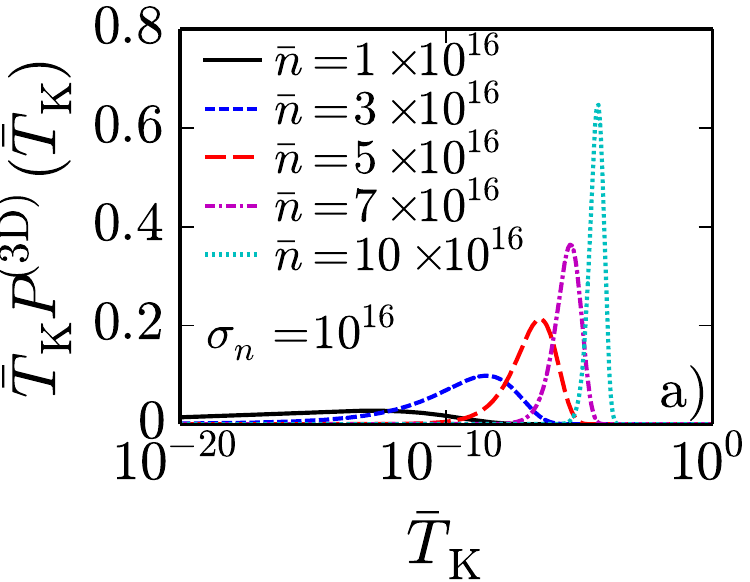}
\includegraphics[width=0.48\columnwidth]{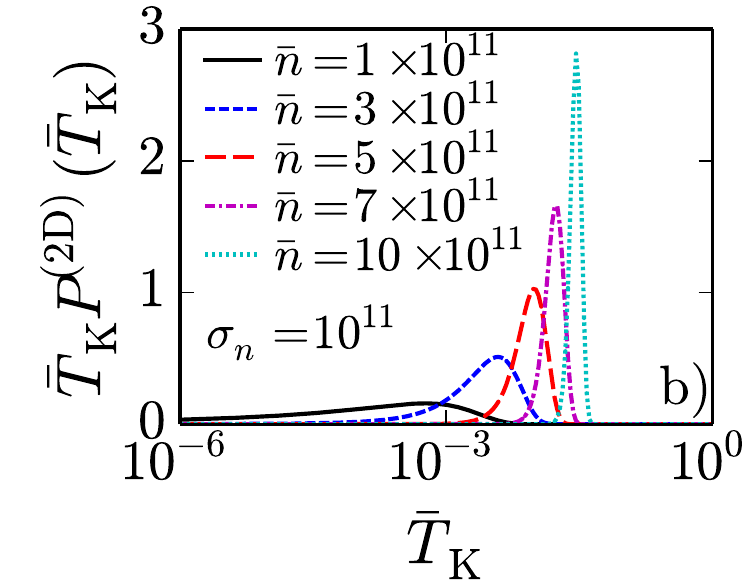}
\end{tabular}
\end{center}
\caption{(Color online).
 ${\bar \tk} P({\bar \tk})$,  ${\bar \tk} \equiv k_{\rm B}\tk/D$, for a 3D, a), and 2D, b), SM  with $v_{\rm F} = 10^8~{\rm cm}/{\rm s}$, $D=0.5$~eV, $J=0.6~J_{\rm cr}$, $N_{\rm w} = 2$  
 and different values of $\bar n$ [in units of ${\rm cm}^{-3}$ in a), ${\rm cm}^{-2}$ in b)].
 In a) $\sigma_n = 10^{16}~{\rm cm}^{-3}$ and $g_s = 1$, in b)  $\sigma_n = 10^{11}~{\rm cm}^{-2}$ and $g_s = 2$.
\label{fig:two}}
\end{figure}

We recall that in 2D $|\mu| \sim n^{1/2}$. 
The major complication in inverting the relation $T_{\rm K}(\mu)$ in this case is due to the asymmetric prefactor $\kappa(\mu)$, which we approximate as $\kappa(\mu) = D$.
In this way we obtain a lower bound for  $P^{({\rm 2D})} (T_{\rm K})$.
\begin{equation} \label{eq:TK_distribution_2D}
P^{({\rm 2D})} (T_{\rm K}) = \frac{\sqrt{2} D^2}{\sqrt{\pi} \sigma_\mu^2 T_{\rm K}} \frac{(1 - J_{\rm c}/J)^2}{|\ln^3(k_{\rm B} T_{\rm K}/D)|} 
\sum_{\lambda=\pm 1} e^{ - \frac{(\mu^2-\lambda {\bar \mu}^2)^2}{2\sigma_\mu^4} } 
~,
\end{equation}
where $\mu \equiv \mu_{\rm 2D}(T_{\rm K})$. 
Eqs.~(\ref{eq:TK_distribution_3D}) and~(\ref{eq:TK_distribution_2D}) show explicitly that, 
in the limit $\tk\to 0$,
\begin{align}
 &P^{({\rm 3D})} (T_{\rm K}) \propto T_{\rm K}^{-1} |\ln(T_{\rm K})|^{-5/2} e^{-{\bar \mu}^6/(2\sigma_\mu^6)}~, \label{eq:ptk3D} \\
 &P^{({\rm 2D})} (T_{\rm K}) \propto T_{\rm K}^{-1} |\ln(T_{\rm K})|^{-3} e^{-{\bar \mu}^4/(2\sigma_\mu^4)}~.   \label{eq:ptk2D}
\end{align}
Our approach is valid as long as the size of Kondo cloud for $T_{\rm K} \geq T$ is smaller
than the correlation length of the disorder potential.
Figure~\ref{fig:two} shows the profile of $P(\tk)$ for different values of $\bar n$ in 3D and 2D, panel a) and b), respectively.
It is interesting to notice that the scaling for $\tk\to 0$  that we find for the 2D case, Eq.~\ceq{eq:ptk2D},
is effectively indistinguishable from the scaling
$P(T_K)\propto T_K^{\alpha -1}$ with $\alpha=0.2$ that was found by fitting NRG results in 
Ref.~\onlinecite{Miranda_prb_2014}.
This result shows the good agreement
in 2D between the NRG  and large-N expansion
and therefore confirms the reliability of the two approaches even in 
the delicate regime induced in 2D Dirac semimetals by the presence of carrier density inhomogeneities.
This agreement also suggests that even in 3D  the large-N expansion should provide
reasonably accurate result for $P(\tk)$.
%

Equations~\ceq{eq:ptk3D}-\ceq{eq:ptk2D} show that in the presence of long-range disorder there is always a large fraction of the sample whose $\tk$ is extremely small. As a consequence at any finite $T$ a significant fraction of carriers is not ``bound'' to the magnetic impurities. From Eqs.~(\ref{eq:TK_distribution_3D}) and~(\ref{eq:TK_distribution_2D}) we determine the number of free spins as $n_{\rm fr}(T) = \int_0^T d T_{\rm K} P (T_{\rm K})$ and in the limit of $T\to 0$ we find
\begin{align}
 &n_{\rm fr}(T) \propto |\ln(T)|^{-3/2}e^{-{\bar n}^2/(2\sigma_n^2)} \hspace{0.25cm}{\rm in~ 3D}~,\label{eq:nfr3D} \\
 &n_{\rm fr}(T) \propto |\ln(T)|^{-2}e^{-{\bar n}^2/(2\sigma_n^2)} \hspace{0.25cm}{\rm in~ 2D} ~.
 \label{eq:nfr}
\end{align}

Note that the number of free spins goes to zero logarithmically as $T\to 0$. Therefore the magnetic susceptibility $\chi_{\rm m}(T) \propto n_{\rm fr}(T)/T$ diverges at low temperature. 
At odds with the magnetic susceptibility of a normal Fermi liquid, $\chi_{\rm m}(T)$ 
does not converge to any finite value for $T=0$ and, away from $T=0$,  does not scale with $T$
as $\sim 1/T$ (Curie-Weiss law)~\cite{Nozieres_JLTP_1974}. This is a clear signature of the development of a NFL behavior. We observe that in Dirac SMs the divergence of $\chi_m(T)$ is stronger than what was found for metals close to a MIT~\cite{Dobrosavljevic_prl_1992}.
Note also that both the distribution $P (T_{\rm K})$ and the number of free spins contain the factor $\exp\big[-{\bar n}^2/(2\sigma_n^2)\big]$, which encodes the effects of both doping and disorder. If the system is strongly doped ({\it i.e.} if ${\bar n} \gg \sigma_n$), the exponential factor strongly suppresses the NFL behavior. The density fluctuations are indeed too small and the Kondo effect is completely controlled by the average Kondo temperature. In this situation, $\chi_{\rm m}$ 
diverges only at extremely small temperatures. On the contrary, when 
$\sigma_n \gtrsim {\bar n}$, the exponential factor is of order of the unity,  and $n_{\rm fr}$ can be quite large.

\begin{figure}[t]
\begin{center}
\begin{tabular}{c}
\includegraphics[width=0.48\columnwidth]{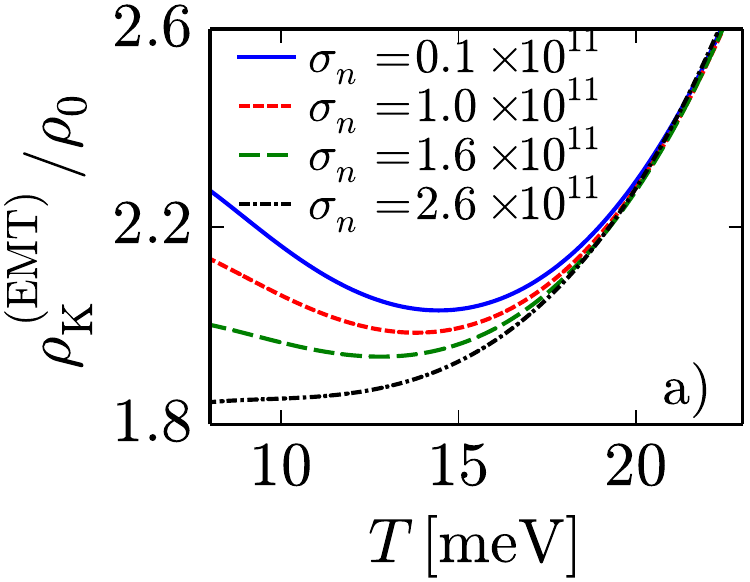}
\includegraphics[width=0.48\columnwidth]{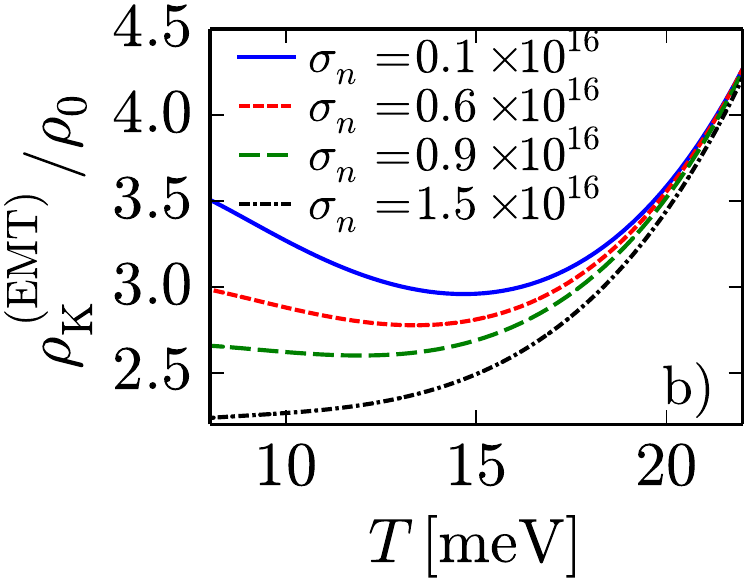}
\\
\includegraphics[width=0.48\columnwidth]{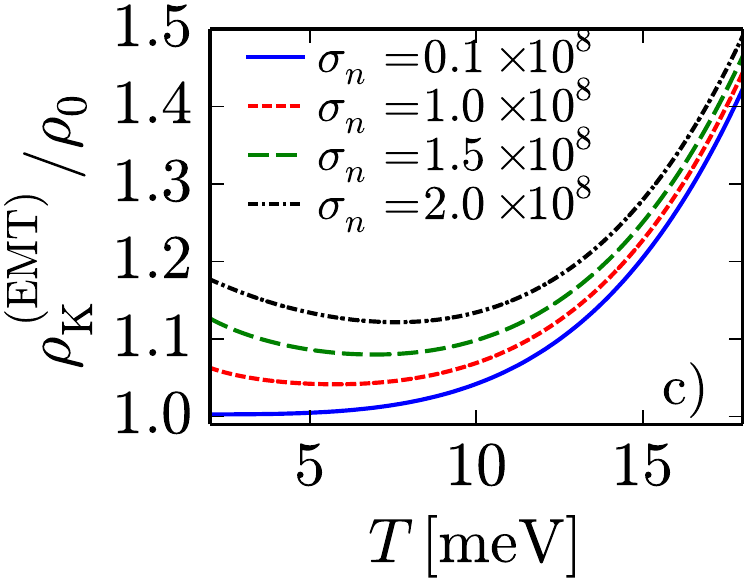}
\includegraphics[width=0.48\columnwidth]{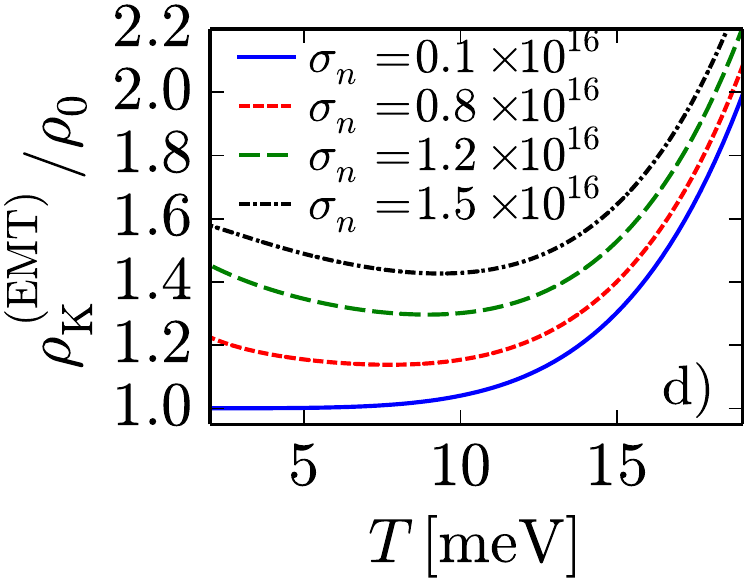}
\end{tabular}
\end{center}
\caption{
(Color online).
          a), b) $\rho_{\rm K}^{({\rm EMT})}(T)$ for the case in which $\bar\mu=60$~meV
          in 2D ($\hat J = 0.98$) and 3D ($\hat J = 1.77$) respectively. 
          c), d) $\rho_{\rm K}^{({\rm EMT})}(T)$ for the case in which $\bar n=0$ ($J< J_{\rm cr}$),
          in 2D and 3D respectively.
          In this plot $\sigma_n$ is in units of ${\rm cm}^{-3}$ in the 3D case, and of ${\rm cm}^{-2}$ in the 2D one.
}
\label{fig:three}
\end{figure}

We now discuss the effect of our results on the transport properties of 3D and 2D Dirac materials.
The coupling term $H_J$ induces a self-energy correction, $\Sigma(\eps)$, for the SM quasiparticles (QPs).  
The imaginary part of $\Sigma(\eps)$ gives the relaxation rate $1/\tau(\eps)$ of the QPs
due their hybridization with the $f$ electrons. 
We find  
$1/\tau(\varepsilon)=4\nimp/\big[\pi {\cal N}(\varepsilon +\mu)\big]$
where $\nimp$ is the density of magnetic impurities.

Notice that $\tau(\varepsilon)$ does not depend on the hybridization $|s|^2$. 
The factor $|s|^2$ due to the interaction vertices between electrons and impurity states is canceled by the opposite factor $\sim 1/|s|^2$
stemming from the spectral weight of impurity states at the Fermi energy.
Using the Boltzmann-transport theory and the expression of $\tau(\varepsilon)$, we can estimate the Kondo resistivity $\rho_{\rm K}$ for the 3D case at $T=0$:
\begin{eqnarray} \label{eq:rho_k}
\rho_{\rm K}(T=0) &=& 
\frac{h}{e^2} \left(\frac{32 g_{\rm s}}{3 \pi^2 N_{\rm w}^2}\right)^{1/3} \frac{n_{\rm imp}}{n^{4/3}}
~.
\end{eqnarray}
It is interesting to compare the scaling given by Eq.~\ceq{eq:rho_k} to that of the resistivity due to {\em scalar} disorder ($\rho$).
for short-range {\em scalar} disorder $\rho$ is independent of $n$~\cite{Burkov_prb_2011}.
For long-range disorder (due to charged impurities) $\rho$~\cite{Burkov_prb_2011}
has the same scaling with respect to $\nimp$ and $n$ as $\rho_{\rm K}(T=0)$. 
The same happens in the 2D case, for which $\rho_{\rm K}=(h/e^2) [4 n_{\rm imp}/(\pi N_{\rm w})] n^{-1}$ ~\cite{Cornaglia_prl_2009}.

To obtain $\rho_{\rm K}(T)$ at finite $T$ it is necessary to keep higher order terms \cite{Hewson_book} in the coupling $J$
and to take into account electron-phonon scattering events. 
We find that in general, within the Bloch-Gr\"uneisen regime for the electron-phonon contribution and for $T>T_{\rm K}$, $\rho_{\rm K}(T)$ is given by the following expression: 
\begin{equation}
\label{eq:rhoT}
\frac{\rho_{\rm K}}{\rho_0} = \left[1 + \frac{1}{4}\frac{(d-1)\pi^2 S(S+1)}{\ln^2(T/T_K) + \pi^2[S(S+1)]^/4} +A_{\rm ph}T^{2+d}\right].
\end{equation}
where $d$ is the dimensionality of the system (2 or 3), $\rho_0 \equiv \rho_{\rm K}(T=0)$ and $\rho_0A_{\rm ph}T^{2+d}$ is the phonon contribution to the resistivity.
This expression is equal to that valid for standard 2D and 3D metallic system.
The unique dispersion of Dirac and Weyl SMs 
affects $\rho(T)$ indirectly through the dependence of $T_K$ on $n$, $J$, and $N_{\rm w}$. Note that in general, also $A_{\rm ph}$ depends on $n$.

The expression of $\rho_{\rm K}$ given by Eq.~\ceq{eq:rhoT} is valid for an homogenous system.
To take into account the effect of the scalar disorder on $\rho_{\rm K}(T)$ we use
the effective medium theory (EMT)
\cite{Bruggeman_annphys_1935,Landauer_JAP_1952,Rossi_prb_2009}. 
In the EMT the resistivity of the inhomogeneous system 
is equal to that of an homogenous ``effective medium'' [$\rho_{\rm K}^{({\rm EMT})}$], and is determined by solving 
the integral equation 
$\int dT_{\rm K} P(T_K) \frac{\rho_{\rm K}^{({\rm EMT})}(T) - \rho_{\rm K}(T,T_{\rm K})}{\rho_{\rm K}^{({\rm EMT})}(T)+(d-1)\rho_{\rm K}(T,T_{\rm K})}=0$. 

In the remainder for the 2D case we use parameter values appropriate for graphene:
$v_{\rm F} = 10^8~{\rm cm}/{\rm s}$, $D=0.5~{\rm eV}$, $N_{\rm w} = 2$, and spin degeneracy $g_s=2$.
In 3D we consider the case of an isotropic linear dispersion
with a Fermi velocity equal to that of graphene, $D=0.5~{\rm eV}$, $N_{\rm w} = 2$ and $g_s =1$,
parameters that roughly approximate the case of ${\rm Cd_2As_3}$~\cite{Neupane_natcomm_2014}.
We then assume ${\hat J} \equiv J{\cal N}(D) = 0.98$ and $A_{\rm ph}=4\times 10^{-6}~{\rm meV}^{-4}$ for the 2D case,
and $\hat J = 0.98$ and $A_{\rm ph}=4\times 10^{-7}~{\rm meV}^{-5}$ for the 3D case.

Figures~\ref{fig:three}a) and~b) show the results for $\rho_{\rm K}^{\rm (EMT)}$, for the 2D and 3D case respectively, when
$\bar\mu=60$~meV ($\bar n=2.647\times 10^{11}~{\rm cm}^{-2}$ in 2D, $\bar n=2.561\times 10^{16}~{\rm cm}^{-3}$ in 3D) and
$\hat J$ is set to a value such that for the homogenous  case we have $T_K=6$~meV [Eq.~\ceq{eq:tk_mu_not_0}], i.e. of the same order
of the values observed experimentally in graphene~\cite{Chen_natphys_2011}. 
We see that for  
$\sigma_n\ll \bar n$, $\rho_{\rm K}^{({\rm EMT})}(T)$ exhibits the nonmonotonic behavior,
characterized by a minimum for $T\sim \tk$, that is the signature of the Kondo effect. 
However for $\sigma_n\gtrsim\bar n$ the profile of $\rho_{\rm K}^{({\rm EMT})}(T)$ changes dramatically:
the minimum  of $\rho_{\rm K}^{({\rm EMT})}(T)$ first becomes shallower, moving to lower values of $T$, and then eventually
disappears.
In both 2D and 3D Dirac SMs, in the presence of long-range disorder,
$\rho_{\rm K}^{({\rm EMT})}(T)$ may not show any qualitative signatures of the Kondo effect
even though in a large fraction of the sample the magnetic impurities are Kondo screened.

We now consider the case in which $\bar\mu=0$. 
Considering that the we have chosen values of  $J< J_{\rm cr}$, in the 
{\em homogenous} limit
$T_K\to 0$ and therefore  $\rho_{\rm K}^{({\rm EMT})}(T)$ does not exhibit any minimum at low $T$.
This picture, however, is qualitatively modified in Dirac materials when long-range scalar
disorder is present, as shown in Figs.\ref{fig:three}~c) and~d):
in the presence of density inhomogeneities, even for $\bar n=0$
and $J<J_{\rm cr}$,  $\rho_{\rm K}^{({\rm EMT})}(T)$ can exhibit a minimum signaling
the presence of Kondo screening in a significant fraction of the sample.




In conclusion, we have studied the Kondo effect in 3D Dirac and Weyl
semimetals. In the absence of long-range, disorder-induced, carrier
density inhomogeneities the Kondo effect is characterized by the Kondo
temperature $\tk$, the crossover temperature below which the Kondo
screening takes effect. When the chemical potential $\mu$ is at
the Dirac point we find that 
no Kondo effect can take place unless the
coupling $J$ between magnetic impurities and conduction electrons is
larger than a critical value $J_{\rm cr}=2/{\cal N}(D)$. In this case
$\tk\propto\sqrt{1-J_{\rm cr}/J}$.
For $\mu>0$, $\tk$ 
depends exponentially on $\mu$ and $J$. 

In the presence of long-range disorder we find that the Kondo effect
is not characterized by a single crossover temperature $\tk$, but by a
distribution of Kondo temperatures $P(T_{\rm K})$.
In the limit
$\tk\to 0$, $P(T_{\rm K})\propto \tk^{-1}|\ln(\tk)|^{-5/2}$ in 3D
and $P(T_{\rm K})\propto T_{\rm K}^{-1} |\ln(T_{\rm K})|^{-3}$ in 2D.
This implies that the magnetic susceptibility diverges
slower than $\sim 1/T$ for $T\to 0$, and that it does not converge to
any finite value at zero temperature, a clear 
signature of a strong
NFL behavior \cite{Dobrosavljevic_prl_1992}.

We have then studied the effect of
Kondo screening, and of the interplay of Kondo screening
and long-range scalar disorder, on the transport properties of Weyl semimetals.
We find that for $T=0$ the Kondo resistivity due to
the presence of magnetic impurities scales as
$\rho_{\rm K}\propto\nimp/n^{4/3}$. We have then obtained the
expression of $\rho_{\rm K}$ for finite $T$ and found that  
when the scalar disorder
is weak $\rho_{\rm K}(T)$ exhibit the typical minimum characteristic of the 
Kondo effect. However, we find that in the presence of strong 
scalar-disorder  $\rho_{\rm K}(T)$ might not show any qualitative signatures of the Kondo effect
even though in a large fraction of the sample the magnetic impurities are Kondo screened
and viceversa exhibit a minimum even in the limit, $\bar n=0$
and $J<J_{c}$, when in the {\em homogenous} system no Kondo-effect is present.
%
%

This work was supported by DOE grant DE-FG02-05ER46203 (AP, GV), by
ONR grant ONR-N00014-13-1-0321 (ER), and by a Research Board Grant at
the University of Missouri (AP, GV).

%

%


\end{document}